\documentclass{INTERSPEECH2023}

\usepackage{pbox}
\usepackage{tikz}
\usepackage{pifont}
\usepackage{enumitem}
\usepackage{multirow}
\usepackage{mathtools}
\usepackage{caption,subcaption}

\interspeechcameraready

\title{
    Unlocking Foundation Models for Privacy-Enhancing Speech Understanding: An Early Study on Low Resource Speech Training Leveraging Label-guided Synthetic Speech Content 
}
\name{Tiantian Feng, Digbalay Bose, Xuan Shi, Shrikanth Narayanan}
\address{
  $^1$Signal Analysis and Interpretation Laboratory, University of Southern California, USA
}
\email{tiantiaf@usc.edu}

\begin{document}

\maketitle
 
\begin{abstract}
Automatic Speech Understanding (ASU) leverages the power of deep learning models for accurate interpretation of human speech, leading to a wide range of speech applications that enrich the human experience. However, training a robust ASU model requires the curation of a large number of speech samples, creating risks for privacy breaches.
In this work, we investigate using foundation models to assist privacy-enhancing speech computing. Unlike conventional works focusing primarily on data perturbation or distributed algorithms, our work studies the possibilities of using pre-trained generative models to synthesize speech content as training data with just label guidance.
We show that zero-shot learning with training label-guided synthetic speech content remains a challenging task. On the other hand, our results demonstrate that the model trained with synthetic speech samples provides an effective initialization point for low-resource ASU training. This result reveals the potential to enhance privacy by reducing user data collection but using label-guided synthetic speech content.

\end{abstract}
\noindent\textbf{Index Terms}: speech emotion recognition, spoken language understanding, privacy, synthetic data, foundation model

\section{Introduction}
\label{section:intro}

Speech provides a natural way for us to express ourselves, enabling the effortless exchange of thoughts, emotions, and ideas. Automatic speech understanding (ASU) typically involves interpreting and comprehending human speech, bringing significant benefits in connecting individuals from diverse backgrounds and assisting individuals with speech impairments. Modern ASU systems typically leverage the power of deep learning \cite{lecun2015deep} models for accurate and robust interpretation of human speech, leading to a wide range of speech applications that improve productivity, enhance the quality of life, and enrich the human experience. For example, popular virtual assistants, like Amazon Alexa, Apple Siri, and Google Assistant, equipped with advanced ASU models, empower numerous novel human experiences that substantially increase user satisfaction.

The success of such an ASU system crucially depends on the performance and robustness of the deployed deep learning models. One critical factor in ensuring the reliability of these deep learning models for ASU relies primarily on collecting qualitative training data. Typically, these datasets include large-scale speech samples from various speakers, languages, and acoustic conditions, leading models to learn diverse variations in speech patterns, intonations, and other environmental factors that may affect speech understanding. However, acquiring speech data frequently raises significant concerns regarding data privacy. In addition to relevant information for developing the ASU system, speech data often carry sensitive information about a person, such as individual demographic attributes (e.g., age, gender), states (e.g., health), or biometric fingerprints (e.g., voice print) \cite{feng2023trust}. Therefore, it is essential for ML practitioners to investigate privacy-enhancing approaches for training ASU models.

Data transformation is widely used to prevent privacy inferences in training the ASU model. For example, researchers applied perturbations to suppress private information from the speech data while minimally interfering with the ASU task \cite{pierre22_interspeech, noe21_interspeech, feng2022enhancing, tomashenko2020voiceprivacy, tomashenko2022voiceprivacy}. Recently, Federated Learning (FL) \cite{konevcny2016federated} has become an emerging privacy-enhancing learning algorithm that allows clients to train a model collaboratively without sharing their data. For instance, prior works explored the ability of FL on diverse ASU tasks, including keyword spotting \cite{zhang2023fedaudio}, ASR \cite{dimitriadis2022flute}, and emotion recognition \cite{feng2022semi}. Although numerous efforts have been made in privacy-enhancing research, researchers demonstrated that the above methods could still leak private information in every possible way \cite{nasr2019comprehensive, zhu2019deep,tseng22_interspeech,feng2021attribute}.

Recent advances in foundation models have empowered many evolving technologies, notably generative AI (e.g., ChatGPT and DALL-E-2 \footnote{https://openai.com/}, etc.), enabling the creation of high-fidelity content in formats of images, audio, and natural language based on user-input requirements or prompts. These advances in generative AI also present opportunities for privacy-enhancing computing, as high-quality generated content poses minimum privacy leaks while offering possibilities to serve as training data. For example, the prior study by \cite{he2022synthetic} demonstrated that training with pure synthetic images could perform better than training with authentic images under zero-shot learning settings. Moreover, researchers studying large language models (LLMs) have broadly adopted synthetic text data as self-instructions for fine-tuning \cite{wang2022self}. Finally, \cite{hu2022synt} also report that using synthetic speech samples in speech recognition training can improve ASR performance. However, these synthetic speech data are still generated from an existing text corpus.

\begin{figure*}[t]
    \begin{center}
        \includegraphics[width=0.9\linewidth]{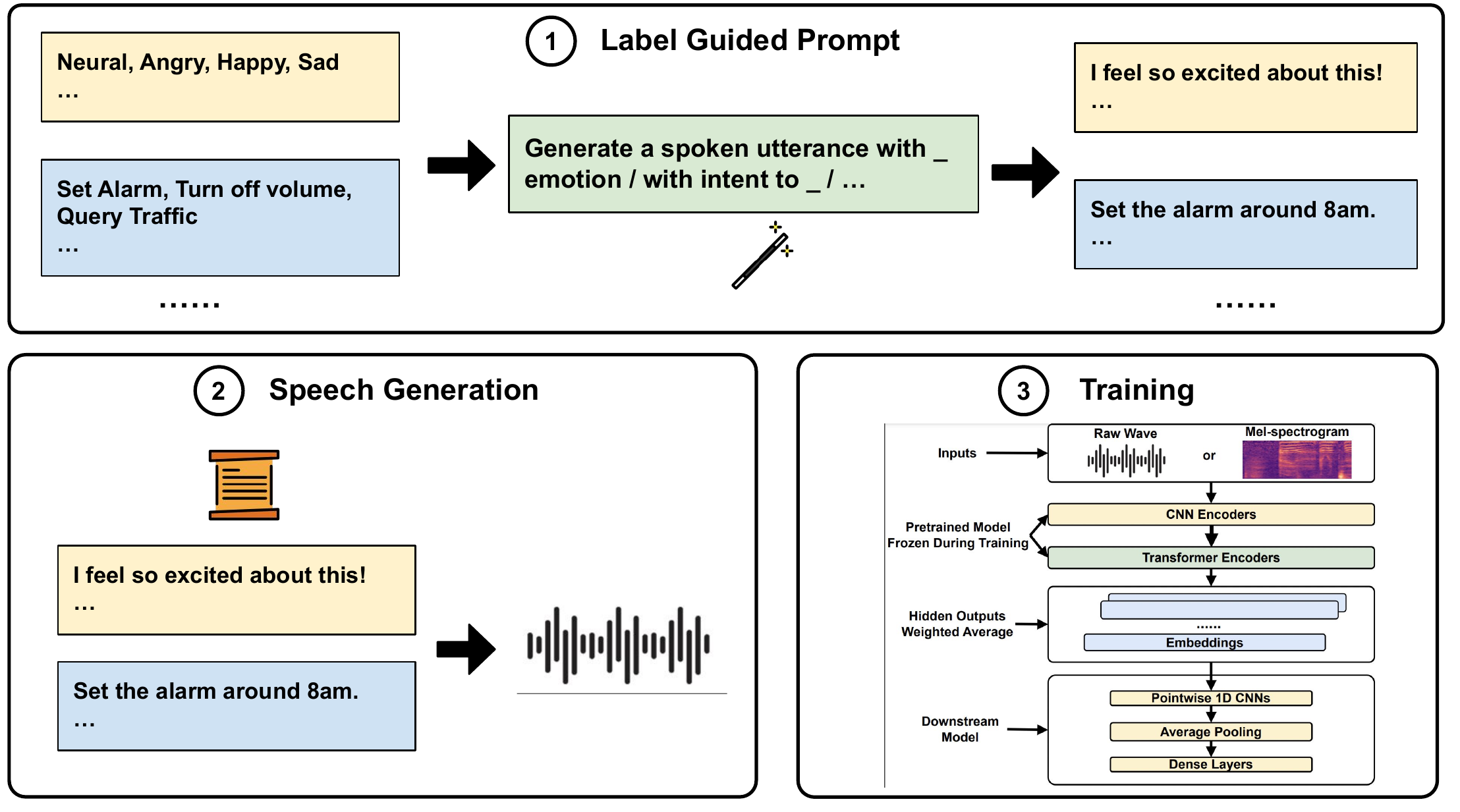}
    \end{center}
    \vspace{-5mm}
    \caption{ASU training framework used for this work leveraging synthetic speech data. The framework begins with label-guided prompts that generate spoken utterances, followed by a text-to-speech model that creates synthetic speech data. The synthetic speech data is used to train the end-to-end ASU model. We use the pre-trained WavLM Base+ as the backbone for fine-tuning.} 
    \label{fig:speech_generation}
    \vspace{-4mm}
\end{figure*}

Inspired by \cite{zhang2023gpt}, in this paper, we provide an early investigation of using label-guided synthetic speech data for ASU training. Unlike previous studies that aim to protect privacy using data transformation or distributed algorithms, \textbf{our method focuses on unlocking the possibilities of using synthetic content as training data to enhance privacy, leading to the reduced need to collect user data or potential in local few-shot training.} Specifically, we perform a two-stage speech synthesis approach combining the LLMs with the text-to-speech (TTS) model. We begin with prompting the LLMs with the label-related information to generate spoken text data which is then transformed into speech data using the text-to-speech (TTS) model. Our proposed method can synthesize speech data without the constraints of an existing text corpus. Specifically, our study aims to answer two fundamental and critical research questions related to privacy-enhancing speech understanding:

\begin{itemize}[leftmargin=*]
    \item Training ASU models with only synthetic data provide the strongest guarantees to user privacy as no actual user data is exposed. However, there is constrained knowledge about the training utilizing label-guided synthetic speech. \textbf{As a result, are label-guided synthetic speech content ready for training ASU models compared to real speech data?}
    
    \item Recent privacy laws, like the EU's General Data Protection Regulation (GDPR) \cite{voigt2017eu}, have prevented the unauthorized collection or misuse of user data, leading to limited data for model training. It is more prevalent to train models with low-resource data to comply with privacy regulations. \textbf{Therefore, are label-guided synthetic speech content ready for low-resource speech training for ASU?}
    
\end{itemize}

\section{ASU Tasks}

This work focuses on two popular ASU tasks: speech emotion recognition (SER) and spoken language understanding (SLU).

\vspace{0.5mm}
\noindent \textbf{Speech Emotion Recognition} \cite{lee2005toward} aims to automatically classify expressed emotions (e.g., neutral, sad, and happy) from speech signals, where the categorical emotion labels are typically acquired from human annotators. Understanding emotions from human conversations has broader applications in virtual assistants, education, and healthcare. 

\vspace{0.5mm}
\noindent \textbf{Spoken Language Understanding} \cite{tur2011spoken} is the task that involves processing speech utterances for domain classification, intent detection, and slot filling. There is a steadily growing interest in deploying robust and efficient SLU systems on mobile and edge applications. Conventional SLU systems typically transcribe the speech signal into spoken format text using ASR models, followed by natural language understanding (NLU) models to recognize the domain and intent. In recent years, researchers have focused on developing end-to-end recipes that directly map speech signals to classify SLU related tasks \cite{serdyuk2018towards}. In this study, we focus on using end-to-end models for SLU tasks.

\section{Two-stage Speech Content Synthesis}

In this work, we propose a two-stage speech synthesis framework as demonstrated in Figure~\ref{fig:speech_generation}. 

\noindent \textbf{Spoken Text Generation} We first apply label information to generate spoken text using LLMs. As shown in Figure~\ref{fig:speech_generation}, the process involves generating spoken utterances that convey particular emotions with prompt messages: Generate a spoken utterance with \_ emotion, where \_ represents categorical emotion labels, such as neutral, happy, and sad. When we aim to generate spoken utterances with intent, we adopt the following prompt message: Generate a spoken utterance with intent to \_. In this study, we use the FLAN-T5 \cite{chung2022scaling} as the language model that consists of 11.3B parameters.

\noindent \textbf{Text-to-Speech} The second stage in speech synthesis involves text-to-speech module. We use the prompted output from spoken text generation as the input to the text-to-speech model to directly synthesize speech data. Specifically, we utilize the recently released SpeechT5 model \cite{ao2021speecht5} for text-to-speech. We augment the speaker information by sampling the x-vector \cite{snyder2018x} from The CMU Arctic dataset \cite{kominek2004cmu}.

\section{Modeling Approach}

\noindent \textbf{Pre-trained Speech Backbone} Our downstream modeling approach utilizes the widely adopted pre-trained speech models, \texttt{WavLM}, as the backbone. WavLM is a self-supervised model that optimizes multiple training objectives, including masked speech prediction, masked speech denoising, and frame prediction. This model is effective in a wide range of ASU tasks \cite{chen2022wavlm}. The WavLM base encoder includes 12 encoding layers with approximately 90M parameters.

\vspace{0.5mm}
\noindent \textbf{End-to-End Downstream Modeling}
Our modeling approach draws inspiration from \cite{pepino21_interspeech}, which highlights that combining the hidden outputs from all encoder layers provides substantially higher performance than relying on the last hidden output for the downstream speech task. In addition, \cite{pepino21_interspeech} also shows that fine-tuning pre-trained speech models with the backbone encoder frozen provides simple but competitive results compared to unfreezing the encoder. In summary, our end-to-end modeling framework starts with weighted averaging to combine the hidden outputs from all encoder layers, where the weights are parameterized. Our downstream model consists of two 1D pointwise convolutional layers with a kernel size of 1 and a filter size of 256. The convolutional layers are connected with the ReLU activation functions. A global average pooling is applied to the convolutional layer output, leading to an output vector of size 256. Following this, the output vector is fed into two fully connected layers for the classification tasks in ASU.

\vspace{0.5mm}
\noindent \textbf{Parameter-Efficient Fine-tuning} In addition to relying on training downstream models, we decide to incorporate \texttt{LoRa (Low-rank Adaptation)} \cite{hu2021lora} in the fine-tuning stage. LoRa adapts the model updates with low-rank matrices during the training phase, bringing benefits to lower inference latency and ease of optimization. Previous work has demonstrated the effectiveness of applying LoRa in fine-tuning ASU tasks \cite{li2023evaluating}.

\begin{table}[t]
    \caption{Summary of dataset statistics used in this work.}
    \vspace{-2.5mm}
    \footnotesize
    \begin{tabular*}{\linewidth}{lccc}
        \toprule
        
        \multirow{1}{*}{\shortstack{\textbf{Datasets}}} & 
        \multirow{1}{*}{\textbf{Unique Speakers}} & 
        \multirow{1}{*}{\shortstack{\textbf{Classes}}} &
        \multirow{1}{*}{\shortstack{\textbf{Total Utterances}}}  \\ 
         
        \midrule
        \textbf{IEMOCAP} & 10 & 4 & 5,531 \\ 
        \textbf{MSP-Improv} & 12 & 4 & 7,798 \\ 
        \midrule
        \textbf{SLURP} & 177 & 46 & 72,277
        \\

        \bottomrule
    \end{tabular*}
    \vspace{-4mm}
    \label{table:datasets}
\end{table}

\section{Datasets}

Table~\ref{table:datasets} displays data statistics for the three datasets included in this work. We use IEMOCAP and MSP-Improv datasets for SER experiments and the SLURP dataset for SLU training.

\vspace{1mm}
\noindent \textbf{Speech Emotion Recognition} \texttt{IEMOCAP} \cite{busso2008iemocap} used in this work contains multi-modal (motion, audio, and video) recordings of acted human interactions from ten subjects, with half males and half females. Moreover, \texttt{MSP-Improv}~\cite{busso2016msp} corpus is developed with the target of investigating naturalistic emotions that were elicited from improvised situations. Similar to the IEMOCAP dataset, the corpus comprises audio and visual data collected from 12 individuals, with an equal number of subjects in both genders. Due to imbalanced label distribution in both datasets, we decided to keep the four most frequently presented emotions for all the datasets, as recommended in \cite{pepino21_interspeech}.

\vspace{1mm}
\noindent \textbf{Spoken Language Understanding} \texttt{SLURP} \cite{bastianelli2020slurp} is the spoken language dataset collected for designing in-home personal robot assistants. The speech data were recorded from over 100 participants reading the collected text prompts carefully crafted by Mechanical Turk (AMT) workers. An example prompt with the intent to ask for time is: How would you ask for the time? Each recording belongs to one unique intent from 46 intent classes. The complete dataset contains a total of 72,277 utterances. We studied the intent classification using the \texttt{SLURP} dataset.

\section{Results}

\subsection{Experiment Details}

\noindent \textbf{Data Split}: We apply 5-fold and 6-fold evaluation on \texttt{IEMOCAP} and \texttt{MSP-Improv} datasets, where each session is regarded as a unique test fold. During each training fold, one data session is used for validation, while the rest are for training. In the \texttt{SLURP} study, we perform the experiment using the standard splits for training, validation, and testing.

\noindent \textbf{Speech Generation}: We generate 1,000 spoken text of each emotion class belonging to neutral, sad, happy, and angry, leading to 4,000 synthetic utterances for SER training. On the other hand, we generate 100 spoken texts with each unique intent associated with the \texttt{SLURP} dataset, resulting in 4,600 synthetic speech samples for training SLU. We control the maximum output token size as 32 and model temperature as 1.0.

\noindent \textbf{End-to-end Training}: We set the batch size as 64 in baseline experiments, including fine-tuning both natural and synthetic data. Specifically, we set the learning rate as 0.0005 and the maximum training epoch as 30 for SER training. We use a learning rate of 0.005 and a maximum training epoch of 50 for SLU training. We set the maximum audio duration as 6 seconds and 3 seconds for SER and SLU datasets, respectively. All experiments are implemented using PyTorch. The experiments are conducted on a high-performance computing server with A40 GPUs. We use the checkpoints of each pre-trained model from HuggingFace \cite{wolf-etal-2020-transformers}.

\begin{figure}[t]
    \begin{center}
        \includegraphics[width=0.85\linewidth]{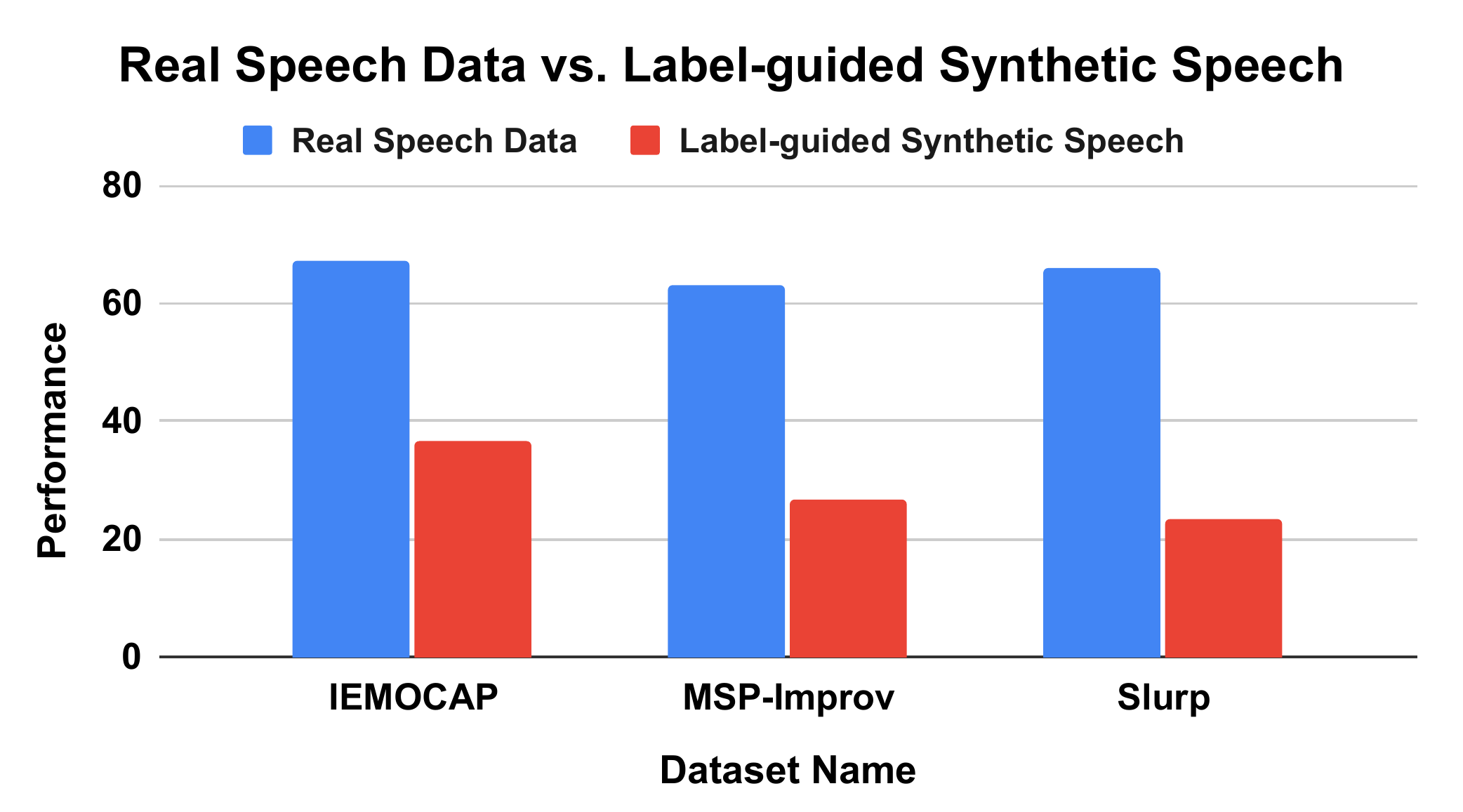}
    \end{center}
    \vspace{-6mm}
    \caption{Baseline fine-tuning performance between real speech data and synthetic speech data.} 
    \label{fig:real_syn_baseline}
    \vspace{-4mm}
\end{figure}

\begin{figure*}[t]
    \begin{center}
        \includegraphics[width=0.87\linewidth]{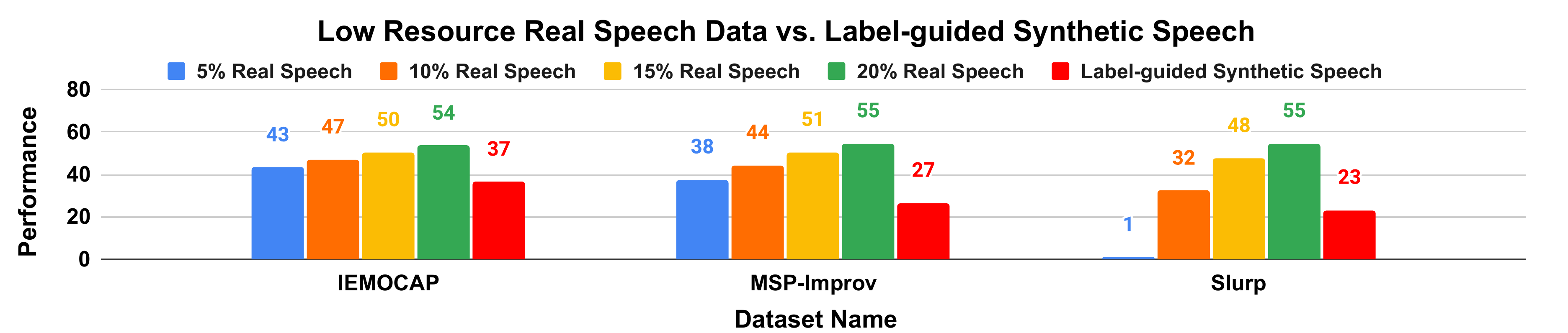}
    \end{center}
    \vspace{-6mm}
    \caption{Fine-tuning performance between low-resource real speech data and synthetic speech data.} 
    \label{fig:low_resource_syn}
    \vspace{-3mm}
\end{figure*}

\begin{figure}[ht] {
    \centering
    
    \begin{tikzpicture}

        \node[draw=none,fill=none] at (0, 5.8){\includegraphics[width=0.925\linewidth]{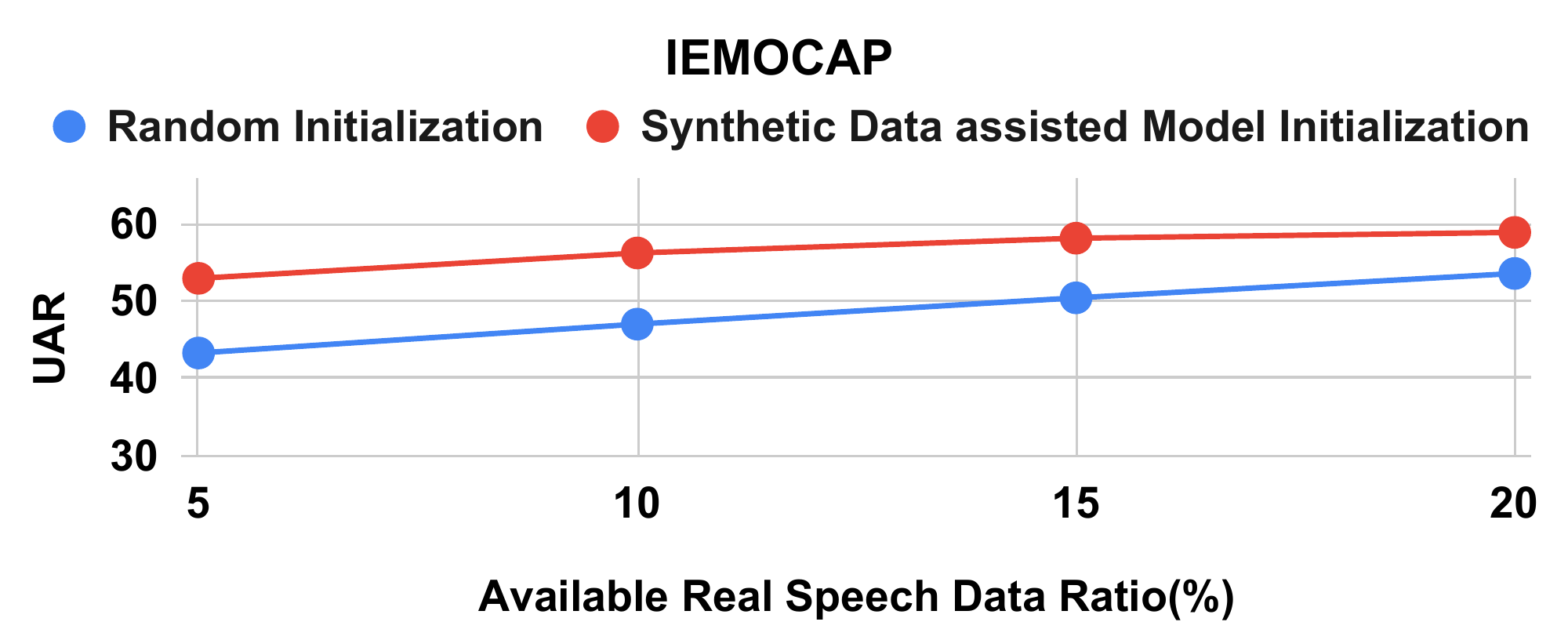}};
        
        \node[draw=none,fill=none] at (0, 2.9){\includegraphics[width=0.925\linewidth]{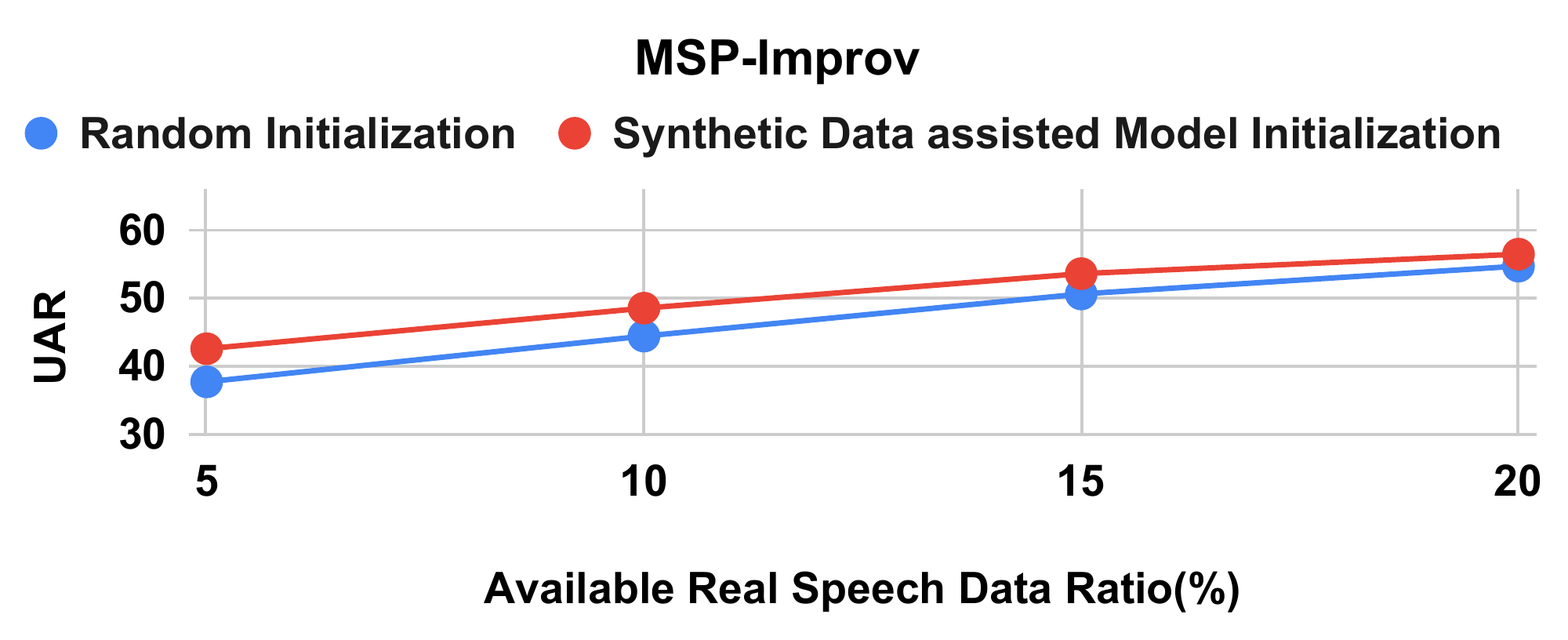}};

        \node[draw=none,fill=none] at (0, 0){\includegraphics[width=0.925\linewidth]{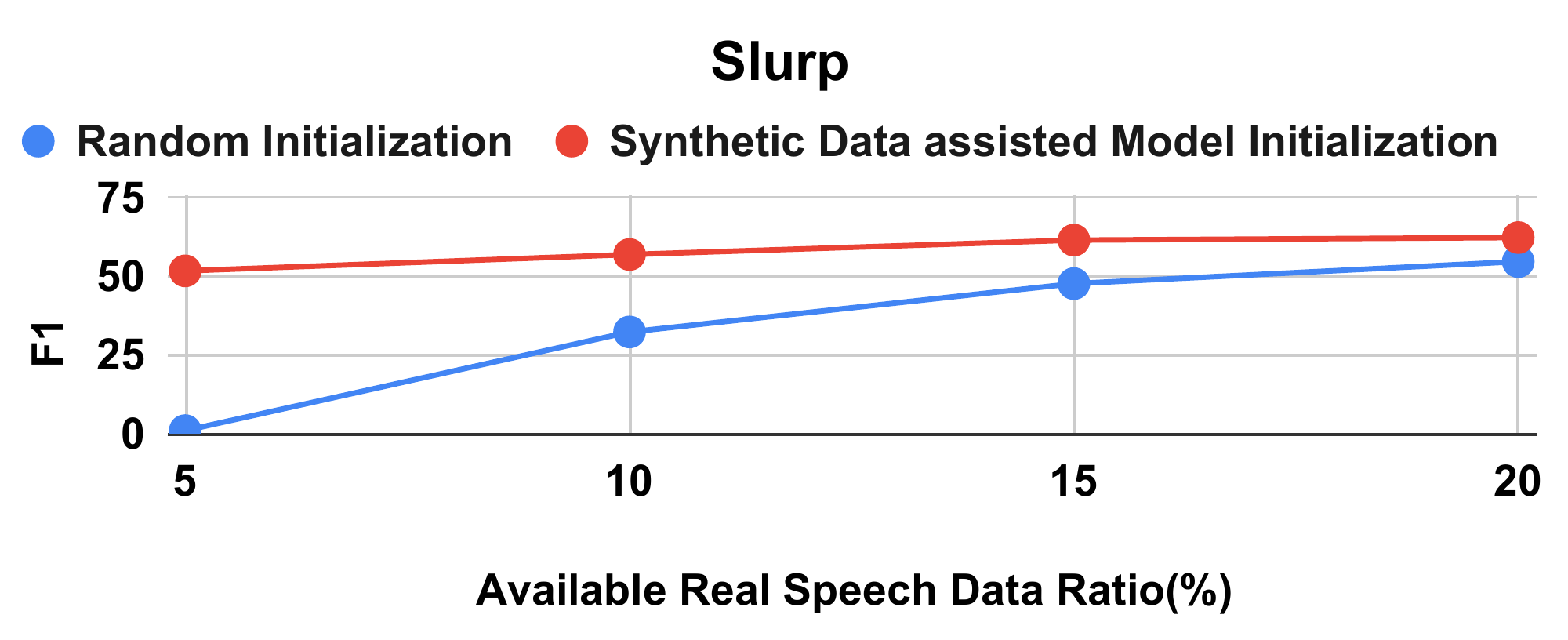}};

    \end{tikzpicture}
    
    \vspace{-3mm}
    \caption{Fine-tuning performance between random initialization and synthetic data assisted model initialization in low-resource speech training. The x-axis represents the available real speech data ratio presented for low-resource training. } 
    
    \label{fig:low_resource_init}

    \vspace{-5mm}
    
} \end{figure}

\subsection{Examples of Spoken Text by Prompt} 

We provide examples of spoken text generated by prompting the LLM to showcase the quality of text input to the TTS module. 

\noindent \textbf{SER Examples} We provide one example of each emotion:

\begin{itemize}[leftmargin=*]
    \item \texttt{Neutral}: "Okay, can you finish a glass of wine, please."
    \item \texttt{Happy}: "We had so much fun in Florida."
    \item \texttt{Sad}: "All I want to do is cry."
    \item \texttt{Angry}: "Why can't you just admit that you broke my car!"
\end{itemize}

\noindent \textbf{SLU Examples} We provide one example with each intent to set alarms, query contact, and mute the volume:

\begin{itemize}[leftmargin=*]
    \item \texttt{Set Alarms}: "You want the alarm to go off in 3 hours."
    \item \texttt{Query contact}: "I'd like to get a list of contacts who are in the San Francisco area."
    \item \texttt{Mute the volume}: "Turn your volume down.",
\end{itemize}

\subsection{Are label-guided synthetic speech content ready for training ASU models?} 

This subsection compares the training performance between real speech data and label-guided synthetic speech data. The results are presented in Figure~\ref{fig:real_syn_baseline}. We use the unweighted average recall (UAR) and Macro-F1 to assess the performance of SER and SLU tasks, respectively. The results demonstrate that zero-shot performance with label-guided synthetic speech data is notably worse than training with real speech data in both SER and SLU tasks. This finding aligns with a prior study \cite{li2018training} that reported a substantial decline in performance when training exclusively with synthetic speech data in ASR tasks. Consequently, it implies that even achieving the same level of performance with real speech training is still challenging using zero-shot learning with label-guided synthetic speech content.

\subsection{Are label-guided synthetic speech content ready for low-resource speech training for ASU?}


Here, we further investigate the role of label-guided synthetic speech content in low-resource speech training. We reduce the learning rate in training SER models to 0.0001 to prevent overfitting from fewer training samples. 

\vspace{0.5mm}
\noindent \textbf{Zero-shot Learning with Synthetic Speech still underperforms Low Resource Speech Training}: We first compare the zero-shot performance with label-guided synthetic speech content and regular training with low-resource speech content. The comparison in Figure~\ref{fig:low_resource_syn} reveals that, in most cases, zero-shot performance with label-guided synthetic speech content fails to provide competitive performance to low-resource speech training. The only exception to this trend is observed when utilizing 5\% real speech with SLURP data for the SLU task. This finding further strengthens the implication that zero-shot learning with label-guided synthetic speech content training is challenging.

\vspace{0.5mm}
\noindent \textbf{Low Resource Speech Training Leveraging Label-guided Synthetic Speech Can Substantially Boost the Performance}: We further investigate low-resource speech training leveraging the label-guided synthetic speech content. Inspired by \cite{zhang2023gpt}, we employ the model trained with label-guided synthetic speech samples as the initialization point for low-resource speech training. Figure~\ref{fig:low_resource_init} illustrates the performance differences between random initialization and synthetic data assisted initialization. The results indicate that, with an equal amount of real training speech data, synthetic data assisted initialization yields a substantial performance improvement compared to random initialization. Moreover, synthetic data assisted initialization leads to comparable ASU performance between training using 20\% and full data (e.g., 20\% data and full data yield 62.32\% and 66.01\% F1 scores on SLURP data, respectively). These findings highlight that although it is challenging to train the ASU model using label-guided synthetic data in lower-resource settings, the synthetic data assisted initialization proves to be effective, significantly enhancing the low-resource training performance.


\section{Implications For Enhancing Privacy}

\noindent \textbf{Privacy-Enhancing Speech Training Leveraging Label-guided Synthetic Speech content is Feasible.} It is promising that under the extremely low-resource condition (5\%), synthetic data assisted model initialization provides competitive model performance. This finding suggests the possibility of collecting significantly fewer private data, therefore reducing privacy concerns and risks associated with data leaks or misuse.

\vspace{0.5mm}
\noindent \textbf{Personalized/Local Learning to Enhance Privacy} The current results also imply the possibility of few-shot learning on the user side without sharing user data. Our future studies would focus on personalized/local learning utilizing synthetic data assisted model initialization to minimize privacy risks.

\vspace{0.5mm}
\noindent \textbf{Improving Text Generation Could Further Enhance the Data-free Zero-Shot Learning} We conjecture that improving the current text prompt would substantially boost the zero-shot learning performance. As presented in Fig~\ref{fig:real_syn_baseline}, the current label-guided text generation may not produce qualitative texts (e.g. low perplexity) for training. This can be related to the constraints of the chosen LLM. These findings emphasize the need for exploring more powerful LLMs such as GPT-4. For example, the current LLM failed to understand the prompt to generate the spoken utterance with the intent to tell a joke, and an incorrect example output is: "Hey man, it's raining again."

\vspace{0.5mm}
\noindent \textbf{Enhance Text-to-speech Quality To Improve Data-free Zero-Shot Learning} Apart from the quality of the text generation, the TTS output also plays a critical role in the zero-shot learning performance with synthetic speech. The existing TTS module lacks the capability to convey a wide range of emotions and prosodic styles comparable to human speech which is essential for generating emotional speech samples. Our future work aims to explore more advanced TTS models.


\bibliographystyle{IEEEtran}
\bibliography{mybib}

\end{document}